\documentclass{PoS}

\usepackage{amsmath}
\usepackage{amssymb}
\usepackage{pstricks}
\usepackage{graphicx}
\usepackage{color}
\usepackage[utf8]{inputenc}

\DeclareMathOperator{\quarkl}{\mathcal{Q}}

\DeclareMathOperator{\smear}{\mathcal{S}}
\DeclareMathOperator{\prop}{K^{-1}}

\title{Towards extracting the timelike pion form factor on CLS 2-flavour ensembles}

\ShortTitle{Towards extracting the timelike pion form factor on CLS 2-flavour ensembles}

\author{\speaker{Felix Erben}\\
        Institut f\"ur Kernphysik and Helmholtz Institut Mainz, Johannes Gutenberg-Universit\"at Mainz, D-55099 Mainz, Germany\\
        E-mail: \email{erben@kph.uni-mainz.de}}

\author{Jeremy Green\thanks{Current affiliation: NIC, DESY, Zeuthen, Germany.}\\
        Institut f\"ur Kernphysik, Johannes Gutenberg-Universit\"at Mainz, D-55099 Mainz, Germany\\
        E-mail: \email{green@kph.uni-mainz.de}}
        
\author{Daniel Mohler\\
        Institut f\"ur Kernphysik and Helmholtz Institut Mainz, Johannes Gutenberg-Universit\"at Mainz, D-55099 Mainz, Germany\\
        E-mail: \email{mohler@kph.uni-mainz.de}}
        
\author{Hartmut Wittig\\
        Institut f\"ur Kernphysik, Johannes Gutenberg-Universit\"at Mainz, D-55099 Mainz, Germany\\
        E-mail: \email{wittig@kph.uni-mainz.de}}

\abstract{Results are presented from an ongoing study of the $\rho$ resonance. We use the distillation approach in order to create correlator matrices involving $\rho$ and $\pi\pi$ interpolators. The study is done in a centre-of-mass frame and several moving frames. We are able to extract energy levels by solving the GEVP of those correlator matrices. The initial exploratory study is being done on a CLS 2-flavour lattice with a pion mass of $451$ $\mathrm{MeV}$ using $\mathcal{O}(a)$ improved Wilson fermions. One aim of this work is to extract the timelike pion form factor after applying the L\"uscher formalism. We also plan to integrate this study with the existing Mainz programme for the calculation of the hadronic vacuum polarization contribution to the muon $g-2$ and will extend our study to lower pion masses and larger lattices in the future, including ensembles with $2+1$ flavours.}

\FullConference{34th annual International Symposium on Lattice Field Theory\\
		24-30 July 2016\\
		University of Southampton, UK}

\begin{document}

\section{Motivation}
The determination of resonance parameters from QCD requires a non-perturbative approach, and lattice QCD is an ideal tool which can help obtaining them. The $\rho$-meson resonance seems to be particularly well-suited for the following reasons: 
\begin{itemize}
\item The noise-to-signal ratio is proportional to $e^{-(m_\rho-m_\pi)}$ (where $m_\pi$ denotes the mass of a pion at rest) which is minimal compared to other hadrons.
\item The $\rho$ almost always decays into a $\pi\pi$ state, which makes it comparably easy to treat on the lattice.
\end{itemize}
It has been shown by Meyer \cite{Meyer:timelike} that one can extract the pion form factor $F_\pi$ in the timelike region $2 m_\pi \leq \sqrt{s} \leq 4 m_\pi$, by computing scattering phase shifts and matrix elements in the vector channel. This quantity is of particular interest to us because it is crucial to reduce the uncertainty in theoretical calculations of the anomalous magnetic moment of the muon \cite{a_hvp}.\\
The $\rho$ resonance also imposes a technical challenge on our calculations: The two pions in the $\pi\pi$ state are interacting in a way which makes it necessary to calculate sink-to-sink quark propagators. One approach which is able to handle those is (stochastic) distillation \cite{dist1,dist2} using Laplacian-Heavyside (LapH) smearing, which is used in this work.

\section{Lattice setup}

\subsection{Distillation}

The inversion of the Dirac matrix $K$ is computationally very expensive. In order to reduce this cost, distillation uses a special kind of hermitian smearing matrix $\smear = V_S V_S^\dagger$ \cite{dist1}. This leads to a much smaller matrix, $V_S^\dagger K^{-1} V_S$, which has to be calculated and stored on disk. Within distillation, quark lines $\quarkl$ (a smeared-to-smeared quark propagator) can be expressed as an expectation value of distillation-sink vectors $\varphi$ and distillation-source vectors $\varrho$:

\begin{align*}
\nonumber \quarkl  &= \sum_b E(\varphi^{[b]}(\rho) ( \varrho^{[b]}(\rho))^\dagger)
,\\
\varphi^{[b]}(\rho) = &\smear \prop V_S P^{(b)} \rho , 
\phantom{aaaa}
\varrho^{[b]}(\rho)  = V_S   P^{(b)} \rho
.
\end{align*}

The noise vectors $\rho$ are defined in the distillation subspace, obeying $E(\rho)=0$ and $E(\rho \rho ^\dagger)=1$. $P^{(b)}$ are the dilution projectors in the distillation subspace.\\
In this work, we use the lowest 56 eigenmodes. Furthermore, we use full distillation \cite{dist1} for the quark lines connected to the source timeslice, and stochastic distillation \cite{dist2} (full spin dilution, 12 dilution vectors in the distillation space and 8 time-dilution vectors) for the sink-to-sink lines. 

\subsection{Lattice ensemble}

This work is an exploratory study on a $\beta=5.3$ CLS 2-flavour lattice. We use $\mathcal{O}(a)$ improved Wilson fermions. The lattice size is $\mathrm{T} \times \mathrm{L}^3 = 64\times32^3$ at a lattice spacing of $a=0.0658 \, \mathrm{fm}$ \cite{params_E5}. The pion mass is $437 \,\mathrm{MeV}$ \cite{params_E5_2}.\\
On this ensemble we analyse 4 different frames: The centre-of-mass frame (CMF) with a total momentum $\mathbf{P} = \frac{2 \pi}{L} \mathbf{d}=0$, as well as three moving frames with lattice frame momenta $\mathbf{d}^2 = {1,2,3}$, averaged over all possible directions on the lattice. In those frames we analyse different lattice irreducible representations (irreps). We use a matrix (the size of which depends on the frame and irrep) which contains a $\rho$ interpolator and $\pi\pi$ interpolators with respective pion momenta $\mathbf{p}_1,\mathbf{p}_2$ obeying $\mathbf{p}_1+\mathbf{p}_2 = \mathbf{P}$:

\begin{align*}
\rho^0(\mathbf{P},t) =
\frac{1}{2 L^{3/2}} \sum_\mathbf{x} e^{-i \mathbf{P} \cdot \mathbf{x}} 
\bigg( \bar{u} (\mathbf{a} \cdot \gamma) u - \bar{d} (\mathbf{a} \cdot \gamma) d \bigg) (t)
\end{align*}
\begin{align*}
(\pi \pi)(t) =
\pi^+(\mathbf{p}_1,t)
\pi^-(\mathbf{p}_2,t)
-\pi^-(\mathbf{p}_1,t)
\pi^+(\mathbf{p}_2,t)
,
\end{align*}
\begin{align*}
\pi^+ (\mathbf{q},t) =
\frac{1}{2 L^{3/2}} \sum_\mathbf{x} e^{-i \mathbf{q} \cdot \mathbf{x}}
\big( \bar{u} \gamma_5 d \big) (\mathbf{x},t),
\phantom{aaa}
\pi^- (\mathbf{q},t) =
\frac{1}{2L^{3/2}} \sum_\mathbf{x} e^{-i \mathbf{q} \cdot \mathbf{x}}
\big( \bar{d}  \gamma_5 u \big) (\mathbf{x},t)
\end{align*}

Since the up- and down-quark masses are degenerate in our calculation ($m_u = m_d$), we work in the isospin limit.

\subsection{Variational method}

For our analysis we use the variational method \cite{gevp1,gevp2}: In this method, in each frame and irrep, we form a correlator matrix $C(t)$ from our interpolators:

\begin{align*}
C(t) = 
\begin{pmatrix}
\langle  \rho(t) \rho^\dagger (0) \rangle  & 
\langle \rho(t) (\pi \pi)^\dagger (0) \rangle \\
\langle  (\pi \pi)(t) \rho^\dagger (0) \rangle &
\langle  (\pi \pi)(t) (\pi \pi)^\dagger (0) \rangle\\
\end{pmatrix}
\end{align*}

\noindent We then solve a generalized eigenvalue problem (GEVP) of this matrix:

\begin{align*}
C(t) \mathbf{v} = \lambda(t) C(t_0) \mathbf{v}
\end{align*}

There are different ways of choosing $t_0$. We use the so-called 'window method' \cite{gevp_window}: $t_0 = t - t_w$. The choice $t_w=3$ gives the best results for our data and is used in this work.\\
The eigenvalues $\lambda$ can be used to extract effective energies because their leading order behaves for large times as $\lambda^{(k)}(t) \rightarrow e^{- E_k \, t_w}$. By taking the logarithm of these eigenvalues, effective energies (effective masses) can be extracted:

\begin{align*}
E_{eff}^{(k)}(t) \equiv \frac{-1}{t_w} \, \ln \lambda^{(k)}(t)
\end{align*}

\section{Analysis}

\subsection{Extracting energy levels}

\begin{figure}
\colorbox{white}{
\includegraphics[width=.99\linewidth]{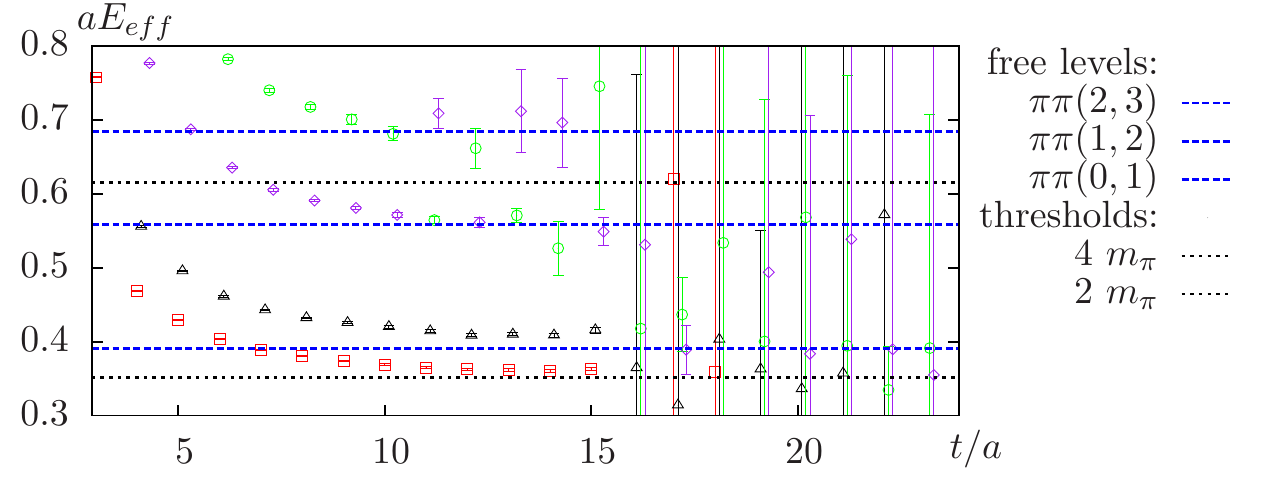}
}
\newline
\begin{flushleft}
\caption{Effective energies of the $4\time 4$-matrix states in the $A_1$ irrep in the $\mathbf{d}^2=1$ moving frame. Each colour depicts an effective energy from a different eigenvalue.}
\label{fig:P1_4}
\end{flushleft}
\end{figure}

\begin{figure}
\colorbox{white}{
\includegraphics[width=.99\linewidth]{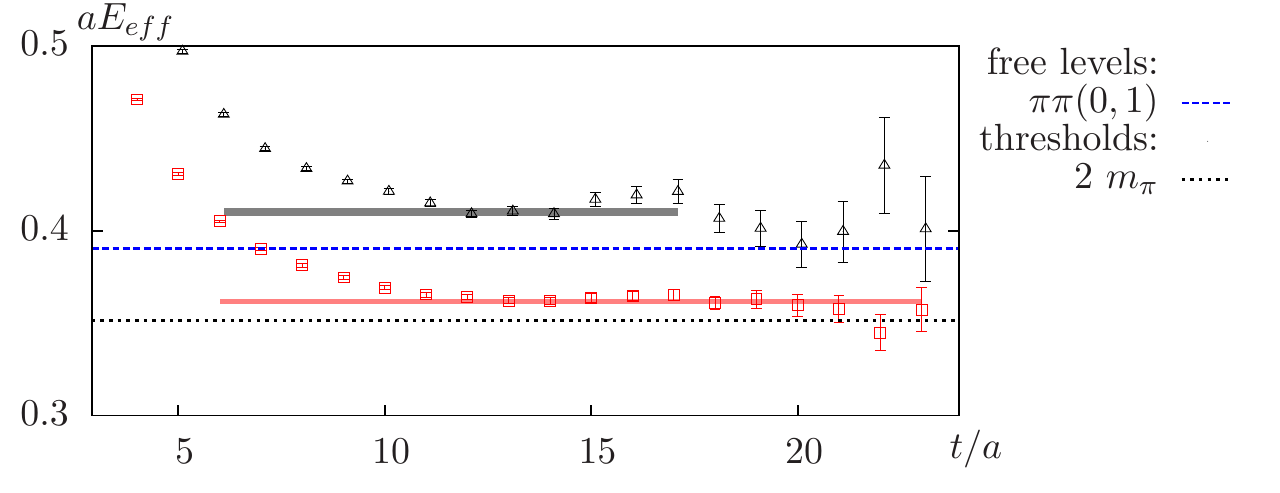}
}
\newline
\begin{flushleft}
\caption{Effective energies of the states in the $A_1$ irrep in the $\mathbf{d}^2=1$ moving frame from a reduced $2\time 2$ matrix with just the lowest $\pi\pi$ interpolator. The bands are fit values from a fit of a constant plus one exponent, fit window depicted by the length of the bands.}
\label{fig:P1_2}
\end{flushleft}
\end{figure}

\begin{figure}
\colorbox{white}{
\includegraphics[width=.99\linewidth]{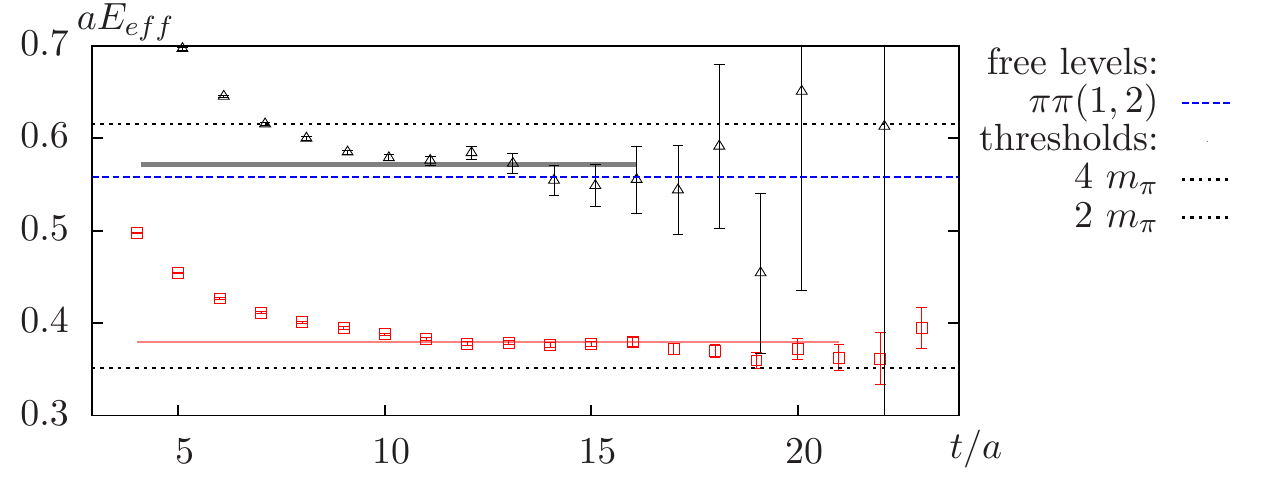}
}
\newline
\begin{flushleft}
\caption{Effective energies of the states in the $E_2$ irrep in the $\mathbf{d}^2=1$ moving frame.}
\label{fig:E2}
\end{flushleft}
\end{figure}

\begin{figure}
\colorbox{white}{
\includegraphics[width=.99\linewidth]{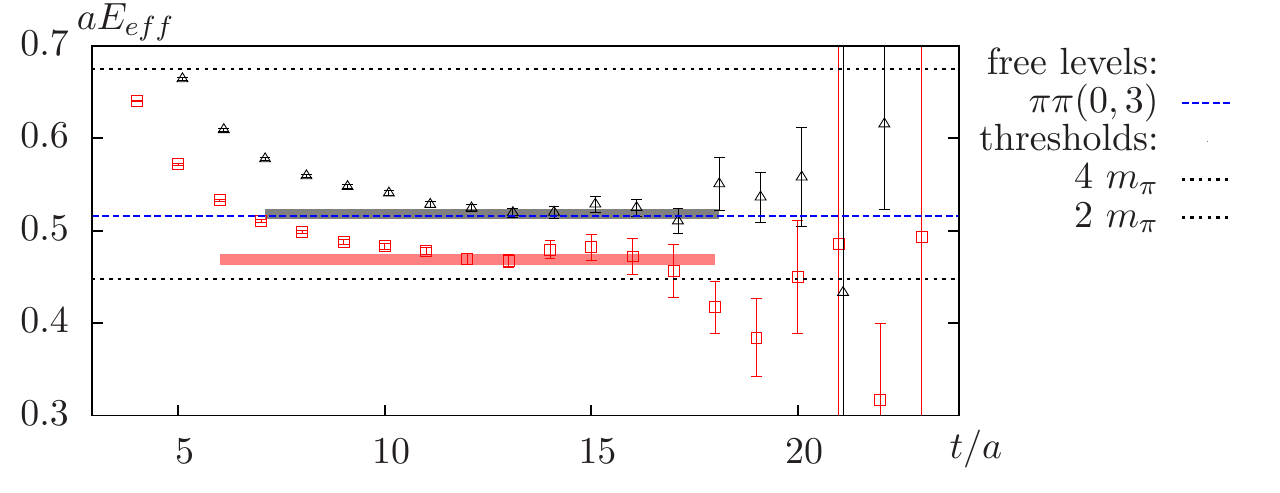}
}
\newline
\begin{flushleft}
\caption{Effective energies of the states in the $A_1$ irrep in the $\mathbf{d}^2=3$ moving frame.}
\label{fig:P3}
\end{flushleft}
\end{figure}

Figure \ref{fig:P1_4} shows the energy levels obtained from a GEVP of the full $4x4$ interpolator matrix in the $A_1$ irrep in the $P^2=1$ moving frame. One can see that we are able to extract the lowest two energy levels with good precision. However, at the current level of statistics, those energy levels can be determined a bit more precisely from the GEVP of a reduced $2x2$ interpolator matrix, using only the lowest of the three $\pi\pi$ interpolators, as can be seen by comparing Figure \ref{fig:P1_4} and \ref{fig:P1_2}. \\
We also used such $2x2$ interpolator matrices in all other irreps and frames. Two examples can be seen in Figure \ref{fig:E2} and \ref{fig:P3}.

\subsection{Phase shift diagram}

L\"uscher's method \cite{luscher1,luscher2} allows us to get information of the continuum spectrum from the energy levels we extracted from the finite-box lattice. It relates the discrete momenta $k$ from the lattice to the continuum phase shift $\delta(k)$ via:

\begin{align*}
\delta(k) + \phi(q) = n \pi
\end{align*}

\begin{figure}
\centering
\colorbox{white}{
\includegraphics[width=.99\linewidth]{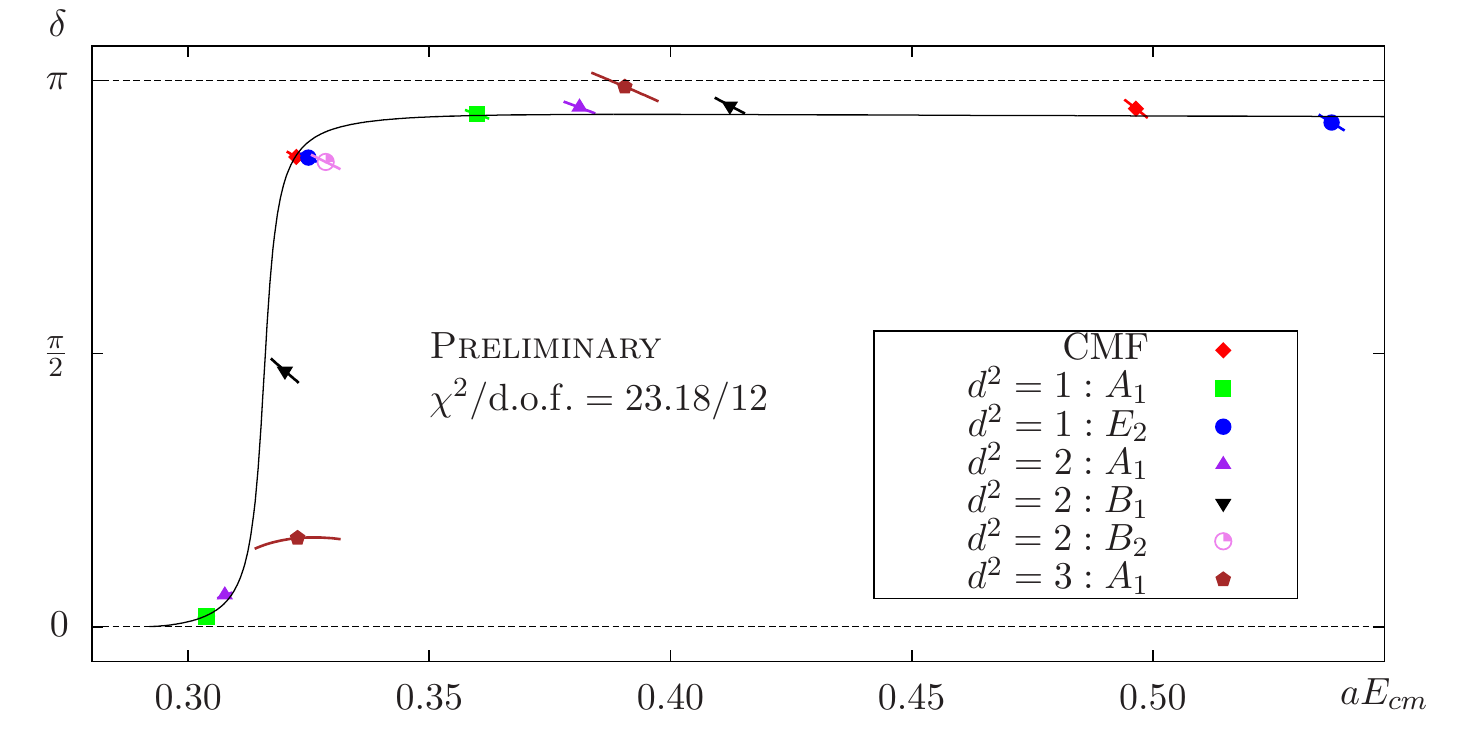}
}
\newline
\begin{flushleft}
\caption{Phase shift $\delta$ plotted as a function of the centre-of-mass energy $E_{cm}$. The different colours correspond to the different irreps in the respective frames. The lowest two energy levels in each irrep have been used for this plot. Because $\delta$ is actually a function of $E_{cm}$ the errorbars in this plot are curvy lines. The correlated fit has $\chi^2 / \mathrm{d.o.f} = 23.8 / 12$.}
\label{fig:phase}
\end{flushleft}
\end{figure}

In here, $\phi$ can be calculated to arbitrary precision and contains modified $\zeta$-functions. The data points in Figure \ref{fig:phase} are determined from those energy levels. For a Breit-Wigner resonance the data is also described by an effective range formula:

\begin{align*}
\frac{p_{cm}^3}{E_{cm}}\cot(\delta) = \frac{6 \pi}{g_{\rho\pi\pi}^2} (m_\rho^2-E_{cm}^2)
\end{align*}

\noindent In this way we determine

\begin{align*}
 a m_\rho = 0.3161(14)  \phantom{aaaa} 
 g_{\rho\pi\pi} = 5.601(73)
\end{align*}

The naive $\rho$ mass extracted from the exponential fall-off of the vector correlator on the same ensemble is $a m_{\rho,\mathrm{naive}} = 0.3208(29)$, which agrees at the level of $1.5 \sigma$. 

\section{Outlook}

This exploratory study is performed on 225 out of 1000 configurations (every fourth from 100 to 1000) on an already well decorrelated ensemble. Two source times are used on each configuration; one of them randomly determined, the other one maximally separated of this first one by $T/2 = 32$. Thus, there is still room to increase statistics further. We will also continue this study on more ensembles with $2+1$ flavours and lighter pion masses.\\
Furthermore, we will calculate correlators such as $\langle V_i(t)O_j^\dagger(0) \rangle$ with pointlike and point-split currents at the sink. Using those we will be able to obtain the matrix elements $\langle \Omega|V_i|k \rangle $ (denoting an overlap of the currents with state $k$, $\Omega$ being the vacuum) which can be used to better understand the large-t behaviour of the vector-vector correlator $G(t) = \langle V_i(t) V_i(0) \rangle$ that is used for $a_\mu^{HVP}$. These matrix elements are also necessary to obtain the pion form factor in the timelike region \cite{Meyer:timelike} from our data, which we can use to determine the $O(\alpha^2)$ hadronic contribution to the muon anomalous magnetic moment more precisely. \cite{a_hvp}


\end{document}